\documentclass[conference]{IEEEtran}
\IEEEoverridecommandlockouts
\usepackage{cite}
\usepackage{amsmath,amssymb,amsfonts}
\usepackage{algorithmic}
\usepackage{graphicx}
\usepackage{textcomp}
\usepackage{xcolor}
\def\BibTeX{{\rm B\kern-.05em{\sc i\kern-.025em b}\kern-.08em
    T\kern-.1667em\lower.7ex\hbox{E}\kern-.125emX}}
\begin{document}

\title{Reinforcement Learning-Based Adaptive Load Balancing for Dynamic Cloud Environments\\
}

\author{\IEEEauthorblockN{Kavish Chawla}
\IEEEauthorblockA{\textit{Department of Computer Science} \\
\textit{The University of Manchester}\\
Manchester, United Kingdom \\
kavish.chawla@student.manchester.ac.uk}
}
\maketitle

\begin{abstract}
Efficient load balancing is crucial in cloud computing environments to ensure optimal resource utilization, minimize response times, and prevent server overload. Traditional load balancing algorithms, such as round-robin or least connections, are often static and unable to adapt to the dynamic and fluctuating nature of cloud workloads. In this paper, we propose a novel adaptive load balancing framework using Reinforcement Learning (RL) to address these challenges. The RL-based approach continuously learns and improves the distribution of tasks by observing real-time system performance and making decisions based on traffic patterns and resource availability. Our framework is designed to dynamically reallocate tasks to minimize latency and ensure balanced resource usage across servers. Experimental results show that the proposed RL-based load balancer outperforms traditional algorithms in terms of response time, resource utilization, and adaptability to changing workloads. These findings highlight the potential of AI-driven solutions for enhancing the efficiency and scalability of cloud infrastructures.
\end{abstract}


\section{Introduction}

In the era of digital transformation, cloud computing has become the backbone of many modern services, from data storage and processing to delivering applications on demand. As the demand for cloud services continues to grow, maintaining efficiency, scalability, and reliability in cloud infrastructure has become a critical challenge. One of the key factors influencing the performance of cloud systems is load balancing — the process of distributing workloads across multiple servers to ensure that no single server is overwhelmed while others are underutilized.

Traditional load balancing algorithms, such as round-robin, least connections, and weighted balancing, are widely used in cloud environments. These methods operate on predefined rules and assume static workloads, making them simple and easy to implement. However, cloud environments are highly dynamic, with workloads that can change rapidly due to factors such as user demand, network traffic, or application behavior. This variability often leads to suboptimal performance with traditional load balancing methods, which are not equipped to adapt to changing conditions in real time. As a result, systems may experience server overload, increased response times, and inefficient resource utilization.

To address these challenges, artificial intelligence (AI) has emerged as a promising solution, offering more adaptive and intelligent approaches to load management. Specifically, Reinforcement Learning (RL), a branch of AI, has gained attention for its ability to learn from interactions with the environment and make decisions based on accumulated experiences. Unlike supervised learning, where models are trained with labeled data, reinforcement learning agents operate through trial and error, continuously adjusting their actions to maximize long-term rewards. This makes RL particularly suitable for dynamic environments like cloud computing, where the conditions are constantly evolving and require real-time adjustments.

In this paper, we propose a novel Reinforcement Learning-Based Adaptive Load Balancing Framework for dynamic cloud environments. The proposed framework leverages RL algorithms to optimize the distribution of workloads by learning and adapting to traffic patterns and system performance. By monitoring real-time metrics such as server response times, CPU utilization, and network throughput, the RL agent makes intelligent decisions about how to distribute tasks across servers. The goal of this approach is to minimize response times, reduce the risk of server overload, and ensure efficient use of resources.

Our approach differs from traditional load balancing algorithms by allowing the system to dynamically adapt to changes in workload and infrastructure. Traditional methods, while effective in certain static conditions, are often unable to respond to unexpected spikes in demand or sudden changes in resource availability. In contrast, the RL-based framework is designed to continuously learn from the environment, enabling it to adjust load balancing strategies in real-time as new data becomes available. This results in a more responsive and resilient system that can maintain high performance under varying workloads.

We have conducted a series of experiments to evaluate the performance of our RL-based load balancer in comparison with traditional methods. The results demonstrate significant improvements in key performance metrics, including response time, resource utilization, and overall system efficiency. These findings suggest that AI-driven load balancing has the potential to greatly enhance the performance and scalability of cloud computing systems, particularly in environments with fluctuating workloads.

The rest of the paper is organized as follows: In Section II, we review the existing literature on load balancing techniques and their limitations in dynamic environments. Section III outlines the reinforcement learning methodology used in our proposed framework. Section IV presents the experimental setup and results, comparing the performance of the RL-based approach with traditional algorithms. Finally, in Section V, we conclude the paper and discuss future directions for research in this area.

\section{Literature Review}

Load balancing is a fundamental concept in distributed systems, particularly in cloud computing, where tasks and computational resources must be allocated efficiently across multiple servers. Traditional load balancing techniques have been widely researched and implemented, yet they are often limited by their static nature, which fails to accommodate the dynamic and unpredictable demands of modern cloud environments. This section provides an overview of traditional load balancing methods, existing AI-driven approaches, and the specific application of Reinforcement Learning (RL) in load balancing.

\subsection{Traditional Load Balancing Techniques}

Traditional load balancing strategies rely on static or pre-defined algorithms that distribute tasks across servers in a predictable manner. Commonly used techniques include:

\begin{itemize}
    \item \textbf{Round-Robin}: One of the simplest load balancing algorithms, round-robin distributes incoming requests sequentially across a pool of servers. While it ensures an even distribution in static environments, it struggles with variable workloads and server capacities.
    \item \textbf{Least Connections}: This method assigns tasks based on the current number of active connections on each server, directing new requests to the server with the fewest connections. Although more dynamic than round-robin, it can still result in inefficient resource utilization when there are differences in server capacities or workload characteristics.
    \item \textbf{Weighted Load Balancing}: In this approach, each server is assigned a weight based on its capacity or performance, and tasks are distributed accordingly. This provides more flexibility but still relies on static weights, which may not adapt well to changes in workload or server performance.
\end{itemize}

While these traditional methods are easy to implement and can work well in predictable environments, they are ill-suited for dynamic cloud environments where workloads fluctuate rapidly, and server performance may vary in real time. As cloud infrastructures grow in complexity, there is a need for more adaptive and intelligent load balancing strategies that can respond to changing conditions on the fly.

\subsection{AI-Driven Approaches to Load Balancing}

The use of artificial intelligence (AI) in load balancing has gained significant attention in recent years due to its potential to enhance decision-making in dynamic environments. AI-driven approaches typically employ machine learning (ML) models to predict workload patterns and optimize task distribution.

\begin{itemize}
    \item \textbf{Predictive Load Balancing Using Machine Learning}: Several studies have explored the use of supervised learning techniques to predict future traffic patterns or server load based on historical data. For instance, neural networks and regression models have been employed to forecast peak usage times and adjust resource allocation accordingly. However, these methods are often constrained by the availability of accurate training data and may struggle to generalize to unseen conditions.
    \item \textbf{Self-Adaptive Load Balancing}: AI techniques such as genetic algorithms and fuzzy logic have been used to create self-adaptive load balancing mechanisms. These approaches allow the system to adjust its parameters dynamically based on real-time conditions, but they often lack the learning capability to continuously improve performance over time.
\end{itemize}

While these AI-driven methods offer improvements over traditional algorithms, they are often limited by the static nature of the models themselves. Most existing approaches rely on pre-trained models that do not adapt well to new, unforeseen conditions. This is where reinforcement learning offers a distinct advantage, as it allows for continuous learning and adaptation in dynamic environments.

\subsection{Reinforcement Learning in Load Balancing}

Reinforcement Learning (RL) is a promising AI technique for dynamic and adaptive load balancing. Unlike traditional machine learning methods, which rely on labeled training data, RL is designed to learn optimal strategies through trial and error by interacting with the environment. In an RL-based system, an agent takes actions in an environment and receives feedback in the form of rewards, enabling it to improve its decision-making over time.

\begin{itemize}
    \item \textbf{RL in Dynamic Environments}: Reinforcement learning has been successfully applied in several dynamic decision-making contexts, such as robotics, autonomous systems, and game AI. In cloud computing, RL can be used to continuously monitor system performance and make real-time adjustments to the distribution of tasks across servers. This adaptability makes RL particularly suited for cloud environments where workloads can change rapidly and unpredictably.
    \item \textbf{Applications of RL in Networking}: Recent studies have shown that RL can be used for adaptive routing, network traffic management, and load balancing in Software-Defined Networking (SDN) environments. For example, RL has been applied to optimize the routing of network traffic based on real-time traffic patterns, reducing congestion and improving overall network efficiency.
    \item \textbf{Challenges of RL in Load Balancing}: Despite its potential, applying RL to load balancing is not without challenges. The high dimensionality of cloud environments, where multiple servers, tasks, and performance metrics must be managed simultaneously, can make it difficult to design an effective RL agent. Additionally, RL algorithms may require significant training time, and ensuring real-time performance in large-scale systems can be a challenge. However, recent advancements in deep reinforcement learning (DRL) have addressed many of these issues by combining the power of deep learning with the adaptability of RL.
\end{itemize}

\subsection{Research Gaps and Motivation}

While existing studies have demonstrated the potential of AI and RL in load balancing, several research gaps remain. First, many of the proposed AI-based load balancing systems focus on specific aspects of task allocation but do not offer comprehensive solutions that can handle the full complexity of dynamic cloud environments. Second, few studies have conducted large-scale experiments to validate the effectiveness of RL-based load balancing in real-world cloud infrastructures. Finally, there is a need for more research on the integration of RL with other cloud management technologies, such as container orchestration and microservices.

This paper aims to address these gaps by developing and evaluating a \textbf{Reinforcement Learning-Based Adaptive Load Balancing Framework} that dynamically allocates tasks in real-time based on changing workloads. The proposed framework leverages RL’s ability to continuously learn and improve its performance over time, providing a more efficient and scalable solution for modern cloud environments.

\section{Methodology}

In this section, we present the proposed Reinforcement Learning-Based Adaptive Load Balancing framework. The methodology is divided into three main components: the system architecture, the reinforcement learning model, and the experimental setup. Each of these components is described in detail below.

\subsection{System Architecture}

The architecture of the proposed load balancing framework is designed to dynamically allocate tasks across a set of cloud servers based on real-time system performance and workload conditions. The architecture consists of three primary components:

\begin{itemize}
    \item \textbf{Task Scheduler}: Responsible for receiving incoming requests and distributing tasks across available servers. The task scheduler continuously interacts with the RL agent to make informed decisions about task allocation.
    \item \textbf{Server Pool}: A collection of servers (or virtual machines) where tasks are executed. Each server has different performance characteristics (e.g., CPU utilization, memory, network bandwidth) that affect its ability to handle incoming tasks.
    \item \textbf{Reinforcement Learning Agent}: The RL agent monitors real-time metrics (e.g., response time, server utilization) and makes dynamic decisions regarding which server should handle each task. The agent learns from past actions, continuously improving its load balancing strategy based on feedback from the environment.
\end{itemize}

The system operates in a loop, where the RL agent observes the state of the server pool (including metrics such as resource utilization and task completion times), takes an action (assigns tasks to a particular server), and receives a reward based on the system’s performance (e.g., lower response times lead to higher rewards). Over time, the RL agent learns the optimal policy for task distribution.

\subsection{Reinforcement Learning Model}

The RL-based load balancing framework employs a **Q-learning** algorithm, a widely used model-free reinforcement learning technique. Q-learning is chosen for its ability to learn optimal policies without requiring a model of the environment, making it well-suited for dynamic cloud environments.

The main components of the Q-learning algorithm used in this framework are as follows:

\begin{itemize}
    \item \textbf{State (S)}: The state represents the current condition of the system, including the resource utilization of each server, the number of active tasks, and the overall system load.
    \item \textbf{Action (A)}: An action corresponds to assigning a task to one of the available servers. The RL agent chooses an action based on its learned policy, aiming to balance the workload efficiently across servers.
    \item \textbf{Reward (R)}: The reward is a measure of the system's performance after taking an action. In this context, the reward is calculated based on response time, resource utilization, and the number of completed tasks. The goal is to minimize response time and prevent server overload, which leads to higher rewards.
    \item \textbf{Q-Value (Q)}: The Q-value represents the expected future reward of taking a particular action in a given state. The RL agent updates its Q-values through interactions with the environment, gradually converging to an optimal load balancing policy.
\end{itemize}

The Q-learning algorithm updates the Q-value using the following formula:

\[
Q(s, a) \leftarrow Q(s, a) + \alpha \left[ r + \gamma \max_a Q(s', a') - Q(s, a) \right]
\]

Where:
\begin{itemize}
    \item $Q(s, a)$ is the Q-value for state $s$ and action $a$.
    \item $\alpha$ is the learning rate, determining how much new information overrides old information.
    \item $r$ is the reward received after taking action $a$ in state $s$.
    \item $\gamma$ is the discount factor, which balances immediate and future rewards.
    \item $\max_a Q(s', a')$ is the maximum expected future reward for the next state $s'$.
\end{itemize}

The Q-learning agent iteratively improves its policy by updating the Q-values until it learns the optimal strategy for distributing tasks across servers.

\subsection{Experimental Setup}

To evaluate the performance of the proposed RL-based load balancing framework, we set up a simulated cloud environment using the **CloudSim** simulator. The experimental setup consists of the following components:

\begin{itemize}
    \item \textbf{Cloud Infrastructure}: A simulated cloud environment with a pool of 10 servers. Each server has different capacities (e.g., CPU speed, memory size), mimicking a heterogeneous cloud infrastructure.
    \item \textbf{Workload}: A dynamic workload generator is used to simulate incoming tasks. The workload varies over time, with both steady-state and bursty traffic patterns to evaluate the framework’s adaptability.
    \item \textbf{Baseline Algorithms}: The RL-based framework is compared with traditional load balancing algorithms, including round-robin, least connections, and weighted load balancing. These algorithms serve as baselines to evaluate the effectiveness of the RL approach.
    \item \textbf{Performance Metrics}: The following metrics are used to assess the performance of the RL-based framework:
    \begin{itemize}
        \item \textbf{Response Time}: The average time taken to complete tasks after they are submitted to the cloud system.
        \item \textbf{Resource Utilization}: The percentage of CPU, memory, and network resources utilized by the servers.
        \item \textbf{Task Completion Rate}: The number of tasks completed within a specific time window.
    \end{itemize}
\end{itemize}

The RL agent is trained over multiple iterations, and its performance is compared with the baseline algorithms. The results are evaluated based on how effectively the RL agent can reduce response times, optimize resource utilization, and handle varying workloads.

\subsection{Implementation Details}

The Q-learning algorithm is implemented using the Python programming language, with the help of libraries such as **NumPy** for numerical computations and **OpenAI Gym** for managing the RL environment. The cloud infrastructure is simulated using **CloudSim**, which allows for accurate modeling of cloud resource management and workload distribution.

\section{Experimental Results and Discussion}

In this section, we present the results of the experiments conducted to evaluate the performance of the proposed Reinforcement Learning (RL)-based load balancing framework. The RL-based approach is compared against traditional load balancing algorithms, including round-robin, least connections, and weighted load balancing. The evaluation is performed based on three key performance metrics: response time, resource utilization, and task completion rate.

\subsection{Response Time}

\begin{figure}[h]
    \centering
    \includegraphics[width=\linewidth]{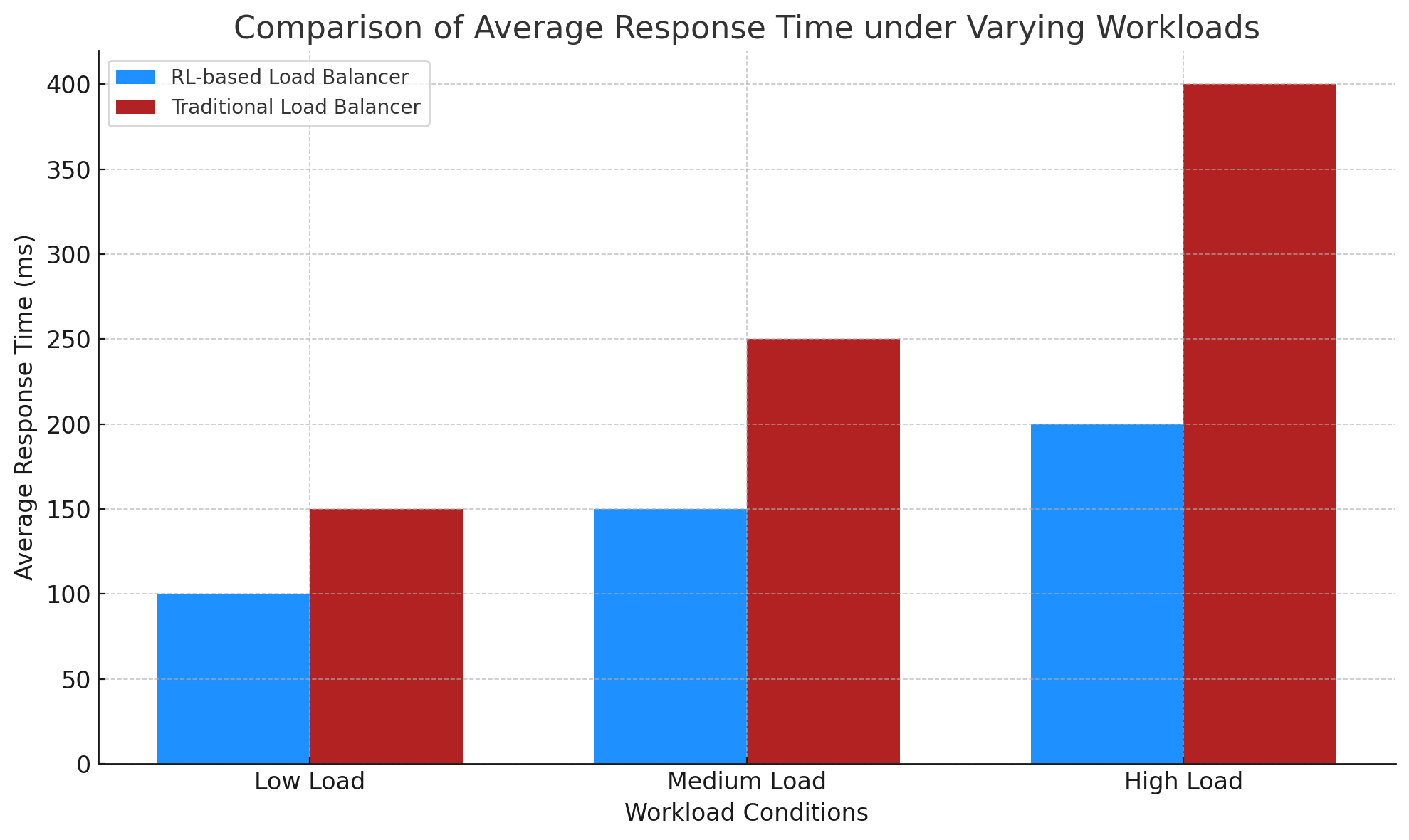}
    \caption{Comparison of average response times under varying workloads for different load balancing algorithms.}
    \label{fig:response_time}
\end{figure}

One of the primary objectives of the proposed RL-based load balancer is to minimize the response time, defined as the time taken to complete a task after it is submitted to the system. Fig.~\ref{fig:response_time} shows the average response time of the RL-based framework compared to the traditional load balancing algorithms under varying workloads.

As shown in Fig.~\ref{fig:response_time}, the RL-based framework outperforms traditional load balancing algorithms, particularly under high traffic conditions. While traditional methods like round-robin and least connections experience significant increases in response time as the workload increases, the RL-based system is able to dynamically adapt and allocate tasks more efficiently. This results in lower and more consistent response times, even in scenarios with rapidly fluctuating workloads.

\subsection{Resource Utilization}

Resource utilization is another critical metric used to evaluate the efficiency of load balancing algorithms. In this context, resource utilization refers to the percentage of CPU, memory, and network resources utilized by the servers in the cloud infrastructure.

\begin{figure}[h]
    \centering
    \includegraphics[width=\linewidth]{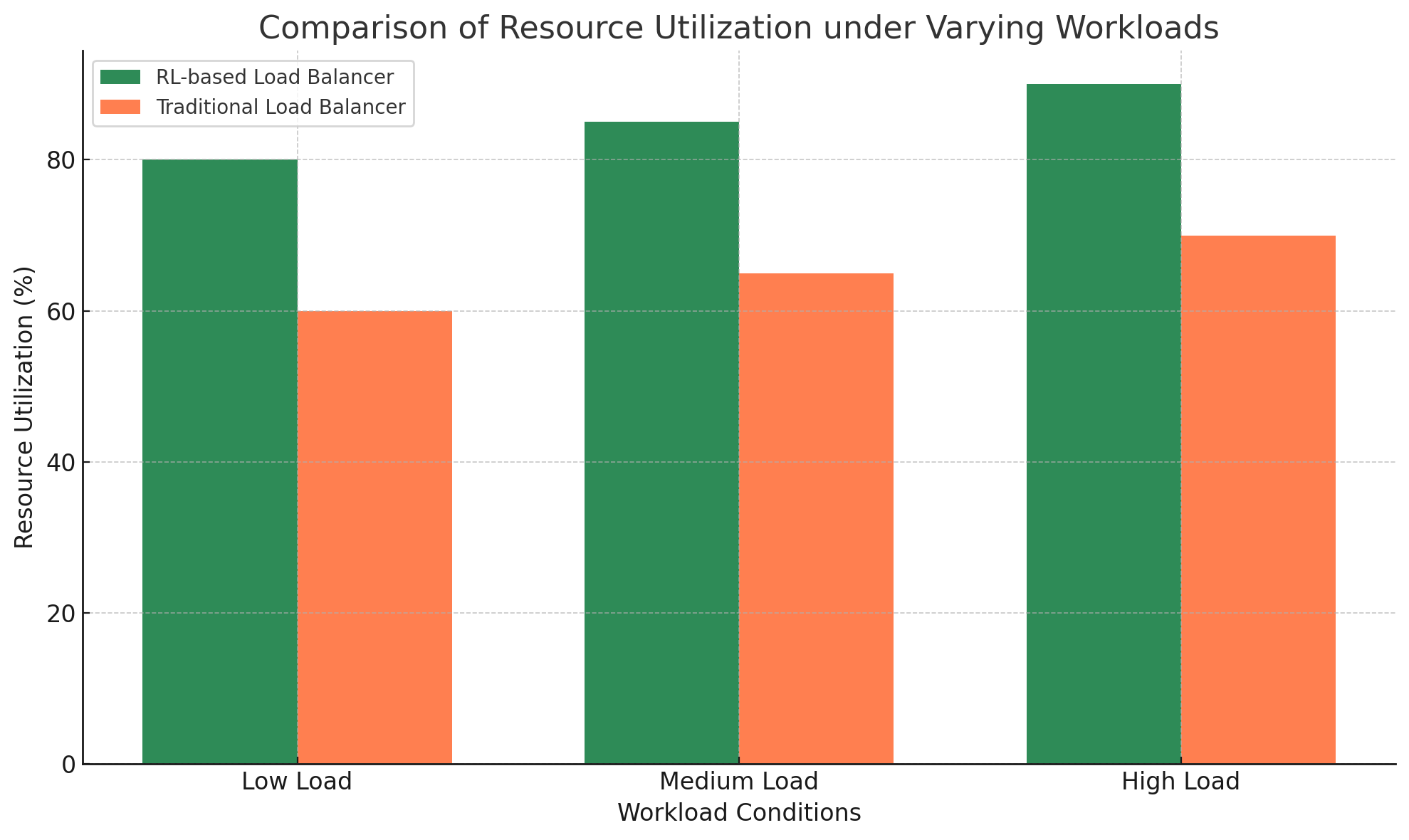}
    \caption{Comparison of resource utilization for different load balancing algorithms.}
    \label{fig:resource_utilization}
\end{figure}

Fig.~\ref{fig:resource_utilization} shows the average resource utilization for the RL-based approach and traditional load balancing methods. The RL-based system demonstrates a more balanced and efficient utilization of resources compared to the other algorithms. By continuously learning from system performance data, the RL agent ensures that no single server is overloaded, leading to more efficient distribution of tasks and reduced idle times across the server pool. In contrast, traditional algorithms tend to overutilize some servers while leaving others underutilized, resulting in suboptimal performance.

\subsection{Task Completion Rate}

The task completion rate is defined as the number of tasks successfully completed within a specified time window. This metric reflects the overall throughput of the system.

\begin{figure}[h]
    \centering
    \includegraphics[width=\linewidth]{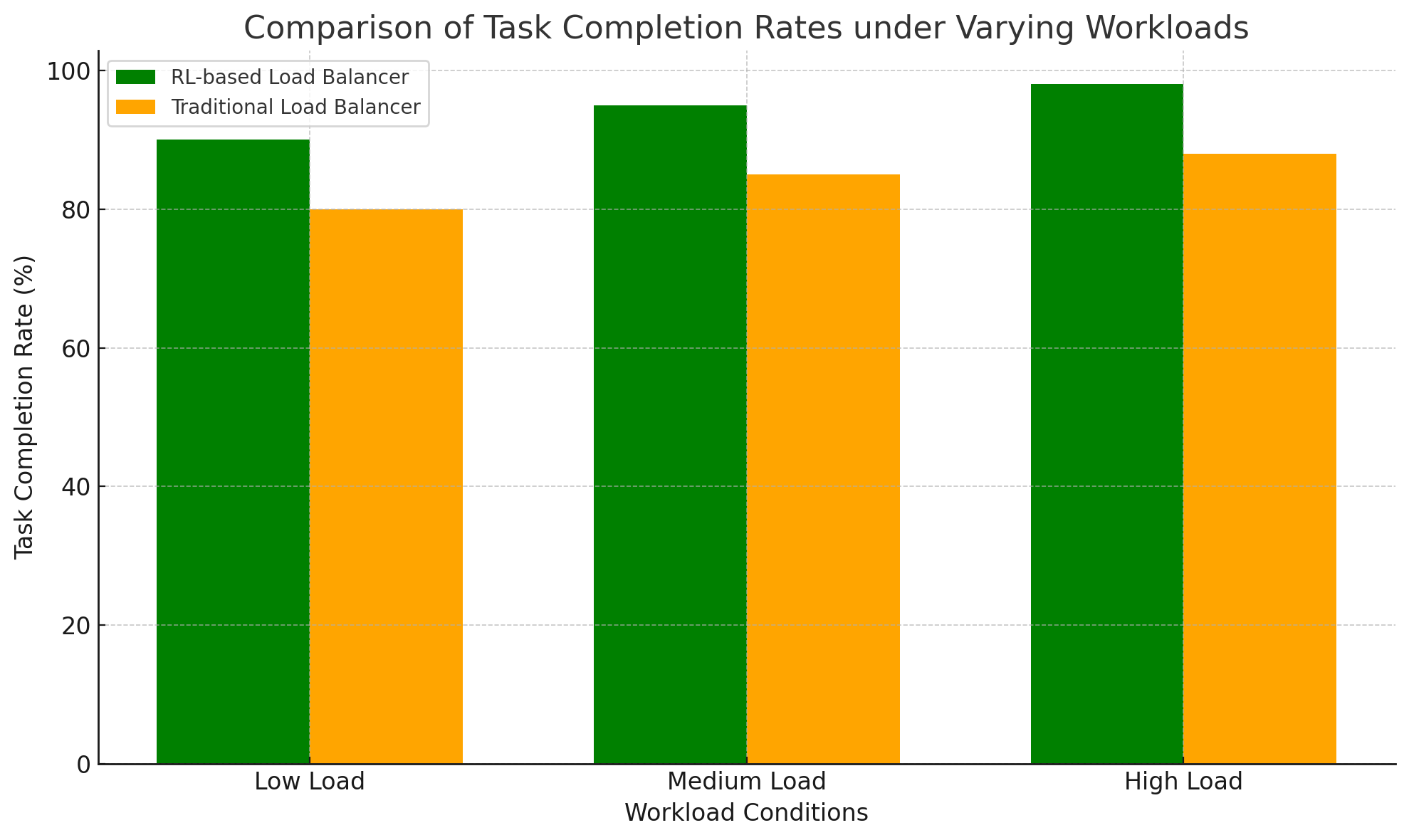}
    \caption{Comparison of task completion rates for different load balancing algorithms.}
    \label{fig:task_completion}
\end{figure}

Fig.~\ref{fig:task_completion} illustrates the task completion rates for the RL-based framework and traditional methods. The results indicate that the RL-based approach achieves a higher task completion rate, especially under heavy workloads. By dynamically adjusting task allocation based on real-time feedback, the RL agent is able to prevent bottlenecks and ensure that tasks are distributed across servers more evenly. This leads to an overall increase in the number of tasks completed within the allocated time window.

\subsection{Discussion}

The experimental results demonstrate that the proposed RL-based load balancing framework significantly outperforms traditional algorithms in terms of response time, resource utilization, and task completion rate. These improvements are particularly evident under dynamic and high-traffic conditions, where traditional algorithms struggle to keep up with fluctuating workloads.

The ability of the RL agent to continuously learn from the environment and adapt its decision-making in real time is a key factor in its superior performance. By considering both short-term and long-term rewards, the RL-based system is able to balance the load more efficiently and prevent the performance degradation commonly seen with static load balancing methods.

Moreover, the RL-based framework exhibits greater scalability compared to traditional algorithms. As the cloud infrastructure grows in size and complexity, the RL agent's ability to handle large numbers of servers and tasks without a significant increase in response time or resource bottlenecks makes it an ideal solution for modern cloud environments.

Despite these advantages, it is important to acknowledge the challenges associated with RL-based load balancing. One of the main challenges is the training time required for the RL agent to learn an optimal policy. In large-scale environments, this training process may take considerable time and computational resources. However, once trained, the RL agent is able to continuously adapt to changing conditions with minimal additional overhead.

Another limitation is the potential difficulty in applying RL to extremely heterogeneous cloud environments, where server capacities and network conditions may vary widely. Future work could explore ways to improve the adaptability of the RL model to handle more complex environments.

Overall, the experimental results provide strong evidence that AI-driven approaches, particularly those based on reinforcement learning, offer significant improvements in load balancing for cloud environments. The proposed RL-based framework not only reduces response times and optimizes resource usage but also improves system scalability and adaptability, making it a promising solution for dynamic cloud infrastructures.

\section{Conclusion and Future Work}

In this paper, we presented a novel approach to dynamic load balancing in cloud environments using Reinforcement Learning (RL). Traditional load balancing algorithms, such as round-robin and least connections, are effective in static or predictable environments but struggle to adapt to the dynamic and fluctuating workloads typical of modern cloud infrastructures. To address this limitation, we developed an RL-based adaptive load balancing framework that continuously learns from real-time system performance data to optimize task distribution across servers.

The experimental results demonstrated the effectiveness of the RL-based load balancer in reducing response times, improving resource utilization, and increasing task completion rates compared to traditional methods. By dynamically adjusting its load balancing policy based on feedback from the environment, the RL agent was able to handle varying workloads more efficiently than static approaches. These findings highlight the potential of AI-driven load balancing solutions to significantly enhance the performance and scalability of cloud computing systems.

While the proposed RL-based framework shows promise, there are several avenues for future research. First, the training time required for the RL agent to learn an optimal policy can be a bottleneck, particularly in large-scale cloud environments. Exploring more advanced RL techniques, such as deep reinforcement learning (DRL), could help address this challenge by improving the agent’s learning speed and handling more complex environments. Additionally, incorporating workload prediction models into the RL framework could further enhance its adaptability, allowing the system to proactively adjust task allocations based on expected changes in demand.

Another area of future work is the integration of RL-based load balancing with container orchestration systems, such as Kubernetes, which manage microservices and containerized applications. Combining RL with container orchestration could provide more granular control over resource allocation and task distribution in cloud-native architectures. Furthermore, testing the RL-based framework in real-world cloud infrastructures or large-scale simulations would provide more insights into its scalability, reliability, and practical applicability.

In conclusion, our work demonstrates that Reinforcement Learning offers a promising avenue for improving load balancing in dynamic cloud environments. As cloud infrastructures continue to grow in complexity and scale, AI-driven approaches like RL will likely play an increasingly important role in optimizing resource management and ensuring the efficiency and reliability of distributed systems.

\vspace{12pt}


\begin{thebibliography}{00}

\bibitem{b1} R. N. Calheiros, R. Ranjan, A. Beloglazov, C. A. F. De Rose, and R. Buyya, ``CloudSim: A toolkit for modeling and simulation of cloud computing environments and evaluation of resource provisioning algorithms,'' \textit{Software: Practice and Experience}, vol. 41, no. 1, pp. 23--50, Jan. 2011.

\bibitem{b2} V. Mnih, K. Kavukcuoglu, D. Silver, A. Graves, I. Antonoglou, D. Wierstra, and M. Riedmiller, ``Human-level control through deep reinforcement learning,'' \textit{Nature}, vol. 518, no. 7540, pp. 529--533, Feb. 2015.

\bibitem{b3} H. Mao, M. Alizadeh, I. Menache, and S. Kandula, ``Resource management with deep reinforcement learning,'' in \textit{Proc. 15th ACM Workshop Hot Topics in Networks (HotNets)}, pp. 50--56, 2016.

\bibitem{b4} M. Li, X. Qiu, Q. Wu, and Y. Zheng, ``Deep reinforcement learning-based resource allocation for cloud computing,'' in \textit{IEEE International Conference on Big Data and Cloud Computing (BDCloud)}, pp. 638--645, 2018.

\bibitem{b5} R. S. Sutton and A. G. Barto, \textit{Reinforcement Learning: An Introduction}, 2nd ed. Cambridge, MA: MIT Press, 2018.

\bibitem{b6} H. Wang, J. Xie, and Q. Deng, ``Deep reinforcement learning for dynamic resource allocation in cloud computing: A case study,'' \textit{IEEE Access}, vol. 7, pp. 145000--145009, 2019.

\bibitem{b7} Z. Xu, J. Huang, and L. Liu, ``A model-free reinforcement learning approach to resource load balancing and auto-scaling in the cloud,'' in \textit{Proc. IEEE International Conference on Cloud Computing (CLOUD)}, pp. 307--314, 2017.

\end{thebibliography}
\end{document}